\documentclass{article}
\usepackage{arxiv}
\usepackage[utf8]{inputenc} %
\usepackage[T1]{fontenc}    %
\usepackage{doi}

\usepackage{graphicx}
\usepackage{amssymb}
\usepackage{amsmath}
\usepackage{amsthm}
\usepackage{booktabs}
\usepackage{todonotes}
\usepackage{adjustbox}
\usepackage{mathtools}
\usepackage{quiver}
\usepackage{array}
\usepackage{verbatim}
\usepackage{hyperref}
\PassOptionsToPackage{hyphens}{url}\usepackage{hyperref}
\usepackage[font=footnotesize,labelfont=bf]{caption}
\usepackage[font=footnotesize,labelfont=bf]{subcaption}
\usepackage[frozencache,cachedir=minted-cache]{minted}

\DeclareMathOperator{\Ob}{Ob}
\DeclareMathOperator{\Hom}{Hom}
\DeclareMathOperator{\src}{src}
\DeclareMathOperator{\tgt}{tgt}

\DeclareMathOperator{\colim}{colim}
\DeclareMathOperator{\op}{op}
\DeclareMathOperator{\co}{co}
\DeclareMathOperator{\Diag}{Diag}
\newcommand{\cat}[1]{\mathcal{#1}}
\newcommand{\C}{$\cat{C}$}
\newcommand{\Set}{\mathbf{Set}}
\newtheorem{theorem}{Theorem}
\newtheorem{proposition}[theorem]{Proposition}

\newcolumntype{C}[1]{>{\centering\let\newline\\\arraybackslash\hspace{0pt}}m{#1}}

\begin{document}
\title{Computational category-theoretic rewriting}

\author{ \href{https://orcid.org/0000-0002-9374-9138}{\includegraphics[scale=0.06]{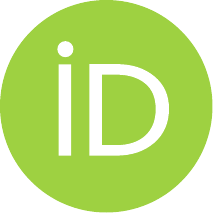}\hspace{1mm}Kristopher Brown} \\
	Topos Institute\\
	\texttt{kris@topos.institute} \\
	\And
	\href{https://orcid.org/0000-0002-8600-949X}{\includegraphics[scale=0.06]{orcid.pdf}\hspace{1mm}Evan Patterson} \\
	Topos Institute\\
	\texttt{evan@topos.institute} \\
	\And 
	Tyler Hanks \\
	University of Florida \\
	\texttt{thanks@ufl.edu} \\
	\And 
	\href{https://orcid.org/0000-0002-1778-3350}{\includegraphics[scale=0.06]{orcid.pdf}\hspace{1mm}James Fairbanks} \\
	Department of Computer Science\\
	University of Florida\\
	\texttt{fairbanksj@ufl.edu} \\
} 
\date{}
\renewcommand{\headeright}{}
\renewcommand{\undertitle}{}

\maketitle 
\begin{abstract}
We demonstrate how category theory provides specifications that can efficiently be implemented via imperative algorithms and apply this to the field of graph rewriting. By examples, we show how this paradigm of software development makes it easy to quickly write correct and performant code. We provide a modern implementation of graph rewriting techniques at the level of abstraction of finitely-presented \C-sets and clarify the connections between \C-sets and the typed graphs supported in existing rewriting software. We emphasize that our open-source library is extensible: by taking new categorical constructions (such as slice categories, structured cospans, and distributed graphs) and relating their limits and colimits to those of their underlying categories, users inherit efficient algorithms for pushout complements and (final) pullback complements. This allows one to perform double-, single-, and sesqui-pushout rewriting over a broad class of data structures.  

\end{abstract}

\section{Introduction and motivation}

Term rewriting is a foundational technique in computer algebra systems, programming language theory, and symbolic approaches to artificial intelligence. While classical term rewriting is concerned with tree-shaped terms in a logical theory, the field of graph rewriting extends these techniques to more general shapes of terms, typically simple graphs, digraphs, multigraphs, or typed graphs. Major areas of graph rewriting are graph {\it languages} (rewriting defines a graph grammar), graph {\it relations} (rewriting is a relation between input and output graphs), and graph {\it transition systems} (rewriting evolves a system in time) \cite{heckel2019analysis}.

When considering the development of software for graph rewriting, it is important to distinguish between studying rewriting systems as mathematical objects and building applications on top of rewriting as infrastructure. The former topic can answer inquiries into confluence, termination, reachability, and whether certain invariants are preserved by rewriting systems. In contrast, we will focus on answering questions that involve the application of concretely specified rewrite systems to particular data. 

Category theory is a powerful tool for developing rewriting software, as the numerous and heterogeneous applications and techniques of rewriting are elegantly unified by categorical concepts. Furthermore, the semantics of categorical treatments of graph rewriting are captured by universal properties of limits and colimits, which are easier to reason about than operational characterizations of rewriting. This is an instance of a broader paradigm of {\it computational applied category theory}, which begins by modeling the domain of interest with category theory, such as using monoidal categories and string diagrams to model processes. One is then free (but not required) to implement the needed categorical structures in a conventional programming language, where the lack of a restrictive type system facilitates a fast software development cycle and enables algorithmic efficiency. For example, arrays can be used to represent finite sets, and union-find data structures can compute equivalence classes. 

Our approach takes the domain of interest modeled by category theory to be the field of graph transformation. This was first suggested by Minas and Schneider \cite{minas2010graph} and is distinguished from existing tools by working at a higher level of abstraction and developing rewriting capabilities within a broader framework of categorical constructions. While current software tools are connected to category theory through their theoretical grounding in adhesive categories \cite{lack2004adhesive}, they are specialized to graphs in their implementation. 

\begin{figure}[h!]
\centering
\includegraphics[width=.7\textwidth]{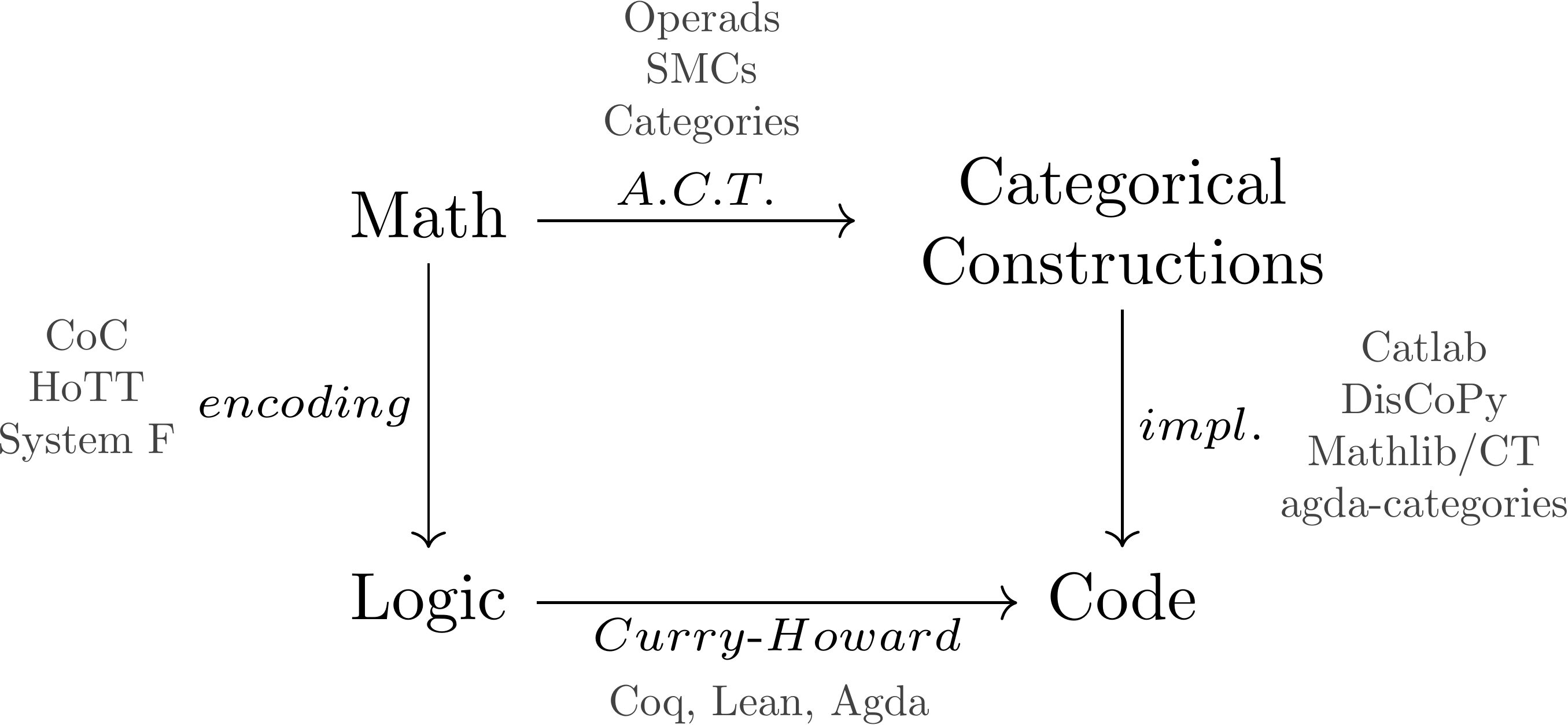}
\caption{Two broad strategies for computational category theory. Applied category theory is used to represent the program's {\it subject matter} in the upper path, while category theory is encoded in the program's {\it structure} or {\it type system} in the lower path. This is not a commutative diagram.}  
\label{fig:software}
\end{figure}

\paragraph{Connection to formal methods}
An orthogonal technique of applying category theory to rewriting software development encodes category theory into the type system of the program itself. This strategy allows type checking to provide static guarantees about the correctness of rewriting constructions. At present, it is not feasible to execute provably-correct programs on large problems, as they generally have poor performance~\cite{ringer2020qed}. Translation-based approaches offer an alternative to proof assistants by encoding graph rewriting into first-order logic and computing answers with SMT solvers, which likewise suffer from scalability concerns when used as an engine to compute rewrites at scale~\cite{heckel2019analysis}. We distinguish computational applied category theory from this paradigm by analogy to the distinction between computational linear algebra and formalizations of linear algebra, a distinction visualized in Figure \ref{fig:software}. One area in which these paradigms can interact is through making the testing of unverified software more robust: extracted programs from formalized proofs can serve as a test oracle and a basis for generating test cases~\cite{rushby2005automated}.

\paragraph{Structure of the paper} 
We will first introduce \C-sets and typed graphs, the latter of which has been the focus of preexisting  graph rewriting software. Our first contribution is to elucidate the subtle relationships between these two mathematical constructs, and we argue on theoretical and performance grounds that \C-sets are more directly applicable to many problems where typed graphs are currently applied. Our next contribution draws from previous theoretical work of L{\"o}we, who developed theory for DPO and SPO of \C-sets \cite{lowe1993algebraic}. We present the first software implementation of this rewriting on \C-sets and extend it with algorithms for SqPO and homomorphism finding. Our last contribution also draws from preexisting theoretical work of Minas and Scheider as mentioned above - we describe a modern realization of computational applied category theory and show how this paradigm allowed for these rewriting techniques to be 1.) efficient, 2.) programmed at a high level, closely matching the mathematical specification, and 3.) extensible to new categories. We lastly outline extensions of rewriting beyond \C-sets, which highlight the flexibility of our technique. 

\section{Important categories in computational graph transformation}
\subsection{Graphs and their homomorphisms}
We take graphs to be finite, directed multigraphs. Thus, a graph $G$ is specified by two finite sets, $G_E$ and $G_V$, giving its edges and vertices, and two functions $G_{\src}, G_{\tgt}:G_E\rightarrow G_V$, defining the source and target vertex of each edge. 

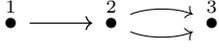
\begin{figure}

\begin{minipage}{.3\textwidth}
\centering
\[
\begin{tikzcd}[]
	{\overset{\tiny 1}{\bullet}} & {\overset{\tiny 2}{\bullet}} & {\overset{\tiny 3}{\bullet}}
	\arrow[from=1-1, to=1-2]
	\arrow[curve={height=-6pt}, from=1-2, to=1-3]
	\arrow[from=1-1, to=1-2]
	\arrow[curve={height=6pt}, from=1-2, to=1-3]
\end{tikzcd}
\]
\end{minipage}
\begin{minipage}{.68\textwidth}
\caption{A graph $G$, defined by $G_V=[3]$, ${G_E=[3]}$, ${G_{\src}=[1,2,2]}$, and ${G_{\tgt}=[2,3,3]}$.} 
\label{fig:grph}
\end{minipage}

\end{figure}

We can compactly represent sets and functions by working in the skeleton of $\mathbf{FinSet}$, where a natural number $n$ is identified with the set $[n] := \{1,...,n\}$. A function $f: [n] \to [m]$ can be compactly written as a list $[x_1,x_2,...,x_n]$, such that $f$ sends the element $i\in [n]$ to the element $x_i \in [m]$. This leads to the edge list representation of graphs, which are encoded as two natural numbers and two lists of natural numbers (Figure \ref{fig:grph}).

Given two graphs $G$ and $H$, a \textit{graph homomorphism} $G\xrightarrow{h} H$ consists of a mapping of edges, $G_E\xrightarrow{h_E} H_E$ and a mapping of vertices, ${G_V \xrightarrow{h_V} H_V}$, that preserve the graph structure, i.e., the following diagrams commute:

\begin{equation}
\label{eq:grhom}
\begin{tikzcd}[]
	{G_E} & {G_V} & {G_E} & {G_V} \\
	{H_E} & {H_V} & {H_E} & {H_V}
	\arrow["{h_E}"', from=1-1, to=2-1]
	\arrow["{h_V}", from=1-2, to=2-2]
	\arrow["{G_{\src}}", from=1-1, to=1-2]
	\arrow["{H_{\src}}"', from=2-1, to=2-2]
	\arrow["{G_{\tgt}}", from=1-3, to=1-4]
	\arrow["{h_V}", from=1-4, to=2-4]
	\arrow["{H_{\tgt}}"', from=2-3, to=2-4]
	\arrow["{h_E}"', from=1-3, to=2-3]
\end{tikzcd}
\end{equation}Regarding the source graph as a pattern, the homomorphism describes a pattern match in the target. A graph homomorphism can also be thought of as a typed graph, in which the vertices and edges of $G$ are assigned types from $H$. For a fixed typing graph $X$, typed graphs and type-preserving graph homomorphisms form a category, namely the slice category $\mathbf{Grph}/X$ \cite{corradini1996graph}.

\subsection{\C-sets and their homomorphisms}

Graphs are a special case of a class of structures called \C-sets.\footnote{\C-sets are also called \emph{copresheaves} on {\C} or \emph{presheaves} on \C$^{op}$, and are what L{\"o}we studied as {\it graph structures} or \emph{unary algebras}.} Consider the category \C~freely generated by the graph $E \overset{s}{\underset{t}{\rightrightarrows}} V$. A \C-set is a functor from the category {\C} to $\mathbf{Set}$, which by definition assigns to each object a set and to each arrow a function from the domain set to the codomain set. For this choice of \C, the category of \C-sets is isomorphic to the category of directed multigraphs. Importantly, we recover the definition of graph homomorphisms between graphs $G$ and $H$ as a natural transformation of functors $G$ and $H$. 

The category \C~is called the \emph{indexing category} or \emph{schema}, and the functor category $[\mathcal{C}, \mathbf{Set}]$ is referred to as \C-$\mathbf{Set}$ or the category of \emph{instances}, \emph{models}, or \emph{databases}. Given a \C-set $X$, the set that $X$ sends a component $c \in \Ob$ \C~to is denoted by $X_c$. Likewise, the finite function $X$ sends a morphism $f \in \Hom_\mathcal{C}(a,b)$ to is denoted by $X_f$. We often restrict to $[\mathcal{C}, \mathbf{FinSet}]$ for computations. %

\begin{figure}[h!]
\centering
\includegraphics[width=\textwidth]{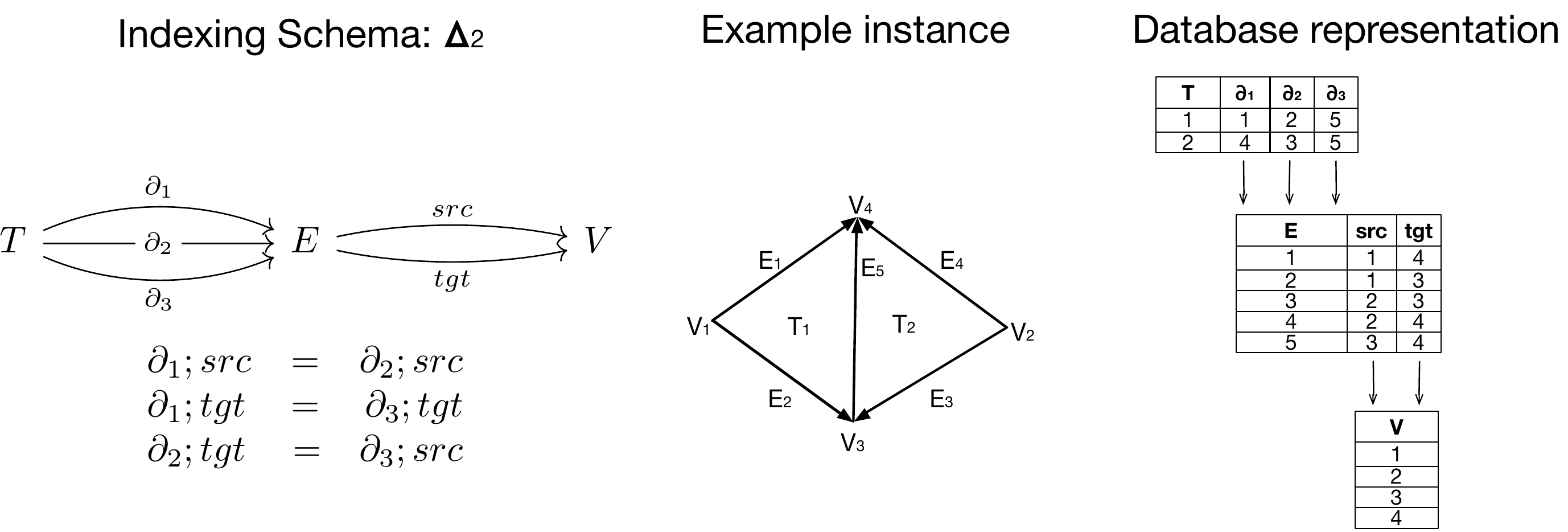}
\caption{The schema of two-dimensional semi-simplicial sets, $\Delta_2$, and an example semi-simplicial set, i.e. an object of $\Delta_2$-{\bf Set}. The equations enforce the connectivity of edges to be a triangle. Note that MacLane defines $\Delta$ as our $\Delta^{op}$.} 
\label{fig:d2}

\end{figure}

In addition to graphs, {\bf Set} itself can be thought of as \C-$\mathbf{Set}$ew, where the schema \C~is the terminal category {\bf 1}. We can change \C~in other ways to obtain new data structures, as illustrated in Figure \ref{fig:d2}. \C-sets can also be extended with a notion of {\it attributes} to incorporate non-combinatorial data \cite{schultz2016algebraic,patterson2021categorical}, such as symbolic labels or real-valued weights. For simplicity of presentation, we focus on \C-sets without attributes in our examples.

\subsection{Relationships between C-sets and typed graphs}

One reason to prefer modeling certain domains using typed graphs or \C-sets rather than graphs is that the domain of interest has regularities that we wish to enforce {\it by construction}, rather than checking that these properties hold of inputs at runtime and verifying that every rewrite rule preserves them. There are close connections but also important differences between modeling with typed graphs or with \C-sets.

\begin{figure}[h!]
\centering
\includegraphics[width=\textwidth]{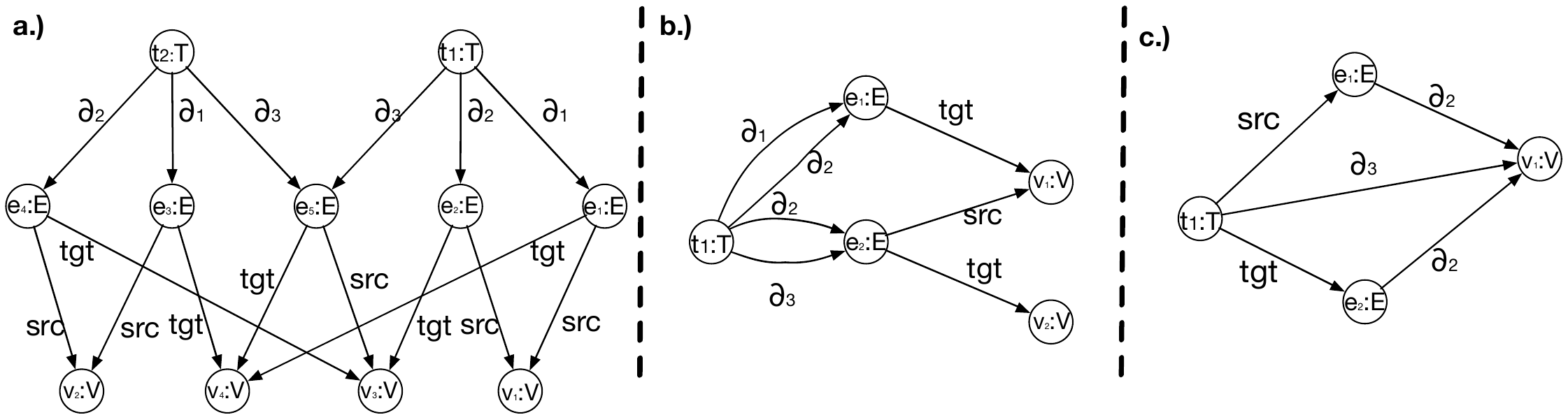}
\caption{{\bf a.)} The semi-simplicial set of Figure \ref{fig:d2}, represented as a typed graph, i.e. a labelled graph with a homomorphism into $\Delta_2$. {\bf b.)} Another valid typed graph which is not a \C-set for three independent reasons: 1.) $T_1$ has multiple edges assigned for $\partial_2$, 2.) $e_1$ has no vertices assigned for $\src$, and 3.) the last equation of $\Delta_2$ is not satisfied. {\bf c.)} A labelled graph which is not well-typed with respect to $\Delta_2$, i.e. no labelled graph homomorphism exists into $\Delta_2$.} 
\label{fig:catelem}
\end{figure}

Every \C-set instance $X$ can be functorially transformed into a typed graph. One first applies the category of elements construction, ${\int X: \mathcal{C}\mathbf{\text{-}Set} \rightarrow \mathbf{Cat}/\mathcal{C}}$, to produce a functor into \C. Then the underlying graph functor $\mathbf{Cat}\rightarrow\mathbf{Grph}$ can be applied to this morphism in {\bf Cat} to produce a graph typed by \C, i.e., a graph homomorphism into the underlying graph of \C. Figure \ref{fig:catelem}a shows a concrete example. However, a graph typed by \C~is only a \C-set under special conditions. The class of \C-typed graphs representable as \C-set instances are those that satisfy the path equations of \C~and are, moreover, \emph{discrete opfibrations} over \C. Discrete opfibrations are defined in full generality in Eq \ref{eq:dof}.\footnote{When specialized to typed graphs, $\mathcal{E} \xrightarrow{F}\mathcal{C}$ is a graph homomorphism and the graphs are regarded as their path categories.}

\vspace{-.4cm} %
\begin{multline}
\text{Given a functor }F: \mathcal{E}\rightarrow \mathcal{C} \text{ : for all } x\xrightarrow{\phi} y \in \Hom \mathcal{C}\text{, and for all } e_x \in F^{-1}(x), \\ \text{there exists a unique } e_x\xrightarrow{e_\phi} e_y \in \Hom \mathcal{E} \text{ such that } F(e_\phi)=\phi
\label{eq:dof}
\end{multline}

\begin{figure}[h!]
\centering
\includegraphics[width=.9\textwidth]{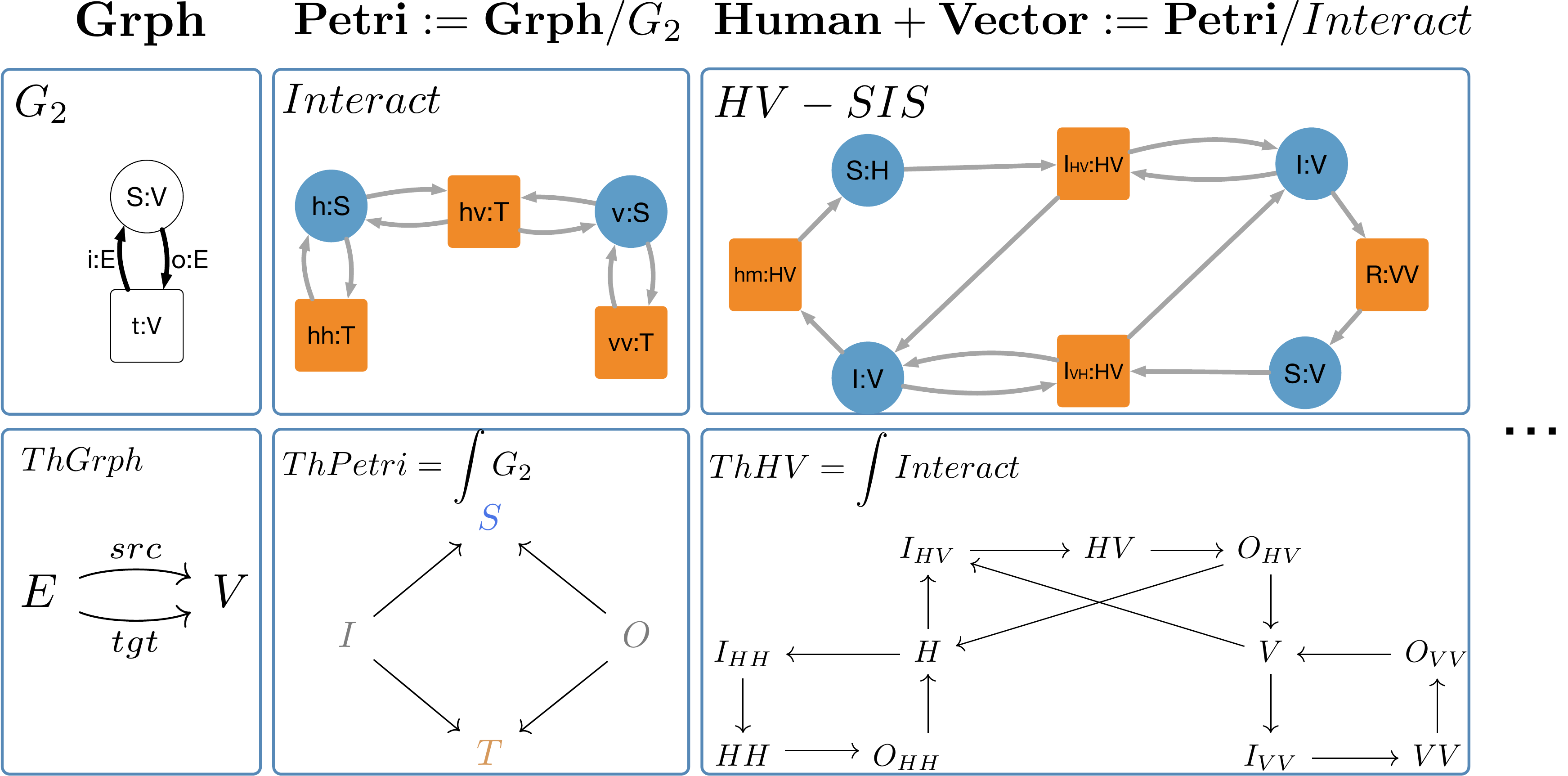}
\caption{Beginning with a theory of graphs, we derive a theory of whole-grain Petri nets (or bipartite graphs) by considering two distinct kinds of vertices (states and transitions) and two kinds of edges (inputs and outputs). $ThPetri$ is constructed the category of elements of $G_2$. Then, taking a slice in {\bf Petri} over an instance, $Interact$, which asserts three kinds of transitions and two kinds of states, we define a type system encoding certain domain knowledge about host-vector interactions, such as the impossibility of a transition which converts a host into a vector. As an example of subtyping, we can interpret hosts as a type of state, implying they are also a type of vertex. This process can be repeated, such as considering SIS disease dynamics for both hosts and vectors. Note that for ease of visualization, \C-set components at the apex of a span of morphisms (e.g. $E$, $I$, $O$) are represented as directed edges. } 
\label{fig:sliceschema}
\end{figure}

However, there is a sense in which every typed graph is a \C-set: there exists a schema $\mathcal{X}$ such that $\mathcal{X}$-{\bf Set} is equivalent to {\bf Grph}$/X$. By the fundamental theorem of presheaf toposes \cite{Kashiwara2006}, $\mathcal{X}$ is the category of elements of the graph $X$, viewed as a $\mathcal{C}$-set on the schema for graphs. Note this procedure of creating a schema to represent objects of a slice category works beyond graphs, which we use to develop a framework of subtype hierarchies for \C-sets, as demonstrated in Figure \ref{fig:sliceschema}.

Because every typed graph category is equivalent to a \C-set category but not the converse, \C-sets are a more general class of structures. The \C-set categories equivalent to typed graph categories are those whose instances represent sets and {\it relations}, in contrast with the general expressive power of \C-sets to represent sets and {\it functions}. Concretely for some edge ${a\xrightarrow{f}b}$ in a type graph $X$, graphs typed over $X$ can have zero, one, or many $f$ edges for each vertex of type $a$, while \C-sets come with a restriction of there being exactly one such edge. While functions can represent relations via spans, the converse is not true.

There are practical consequences for this in graph rewriting software, if one is using typed graph rewriting to model a domain that truly has functional relationships. Because rewrite rules could take one out of the class of discrete opfibrations, as in Figure \ref{fig:catelem}b, this becomes a property that one has to verify of inputs and check all rewrite rules preserve. Typed graph rewriting software can allow declaring these constraints and enforce them, but this becomes an additional engineering task outside of the underlying theory. In contrast, \C-sets are discrete opfibrations by construction.

Path equations are another common means of modeling a domain that are not represented in the theory of typed graph rewriting. This means, for example, that the equation $\partial_1;tgt = \partial_2;src$ in a semi-simplicial set must be checked of all runtime inputs as well as confirmed to be preserved by each rewrite rule. This property is not straightforward to guarantee in the case of sesqui-pushout rewriting. As an upcoming example will demonstrate, it is not sufficient to just check that one's rewrite rule satisfies the path equalities: the rewriting itself must take path equalities into account in order to compute the correct result.

\begin{figure}[h!]
\centering
\includegraphics[width=1\textwidth]{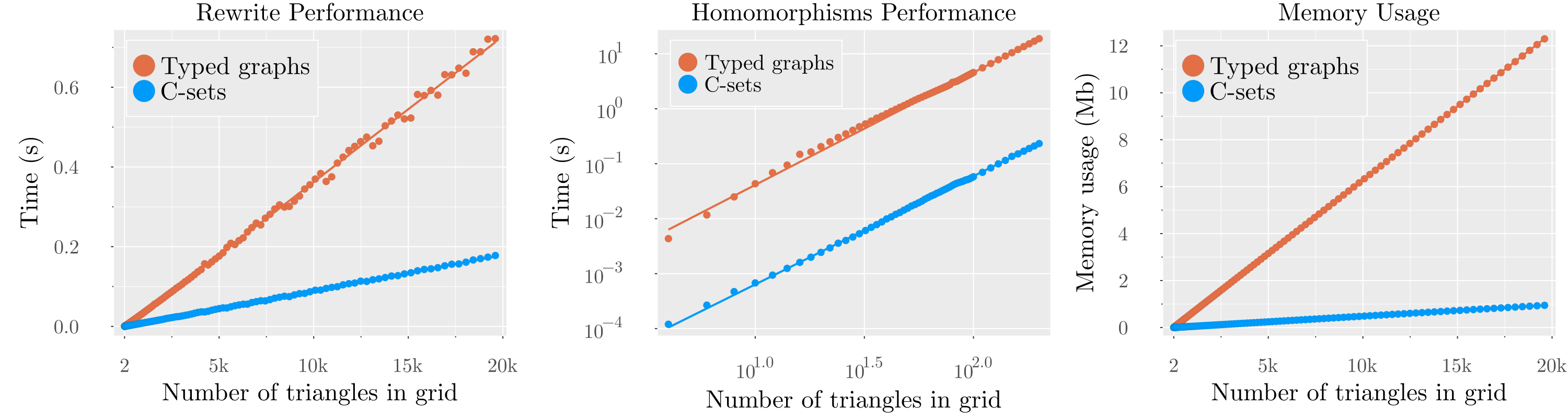}
\caption{Semisimplicial set edge flip benchmark results.  Time was measured on an AMD EPYC 75F3 Milan 3.0 GHz Core with 4GB of allocated RAM.} 
\label{fig:intbench}
\end{figure}

Furthermore, there are performance improvements made possible by working with \C-sets, rather than typed graphs. Borrowing terminology from relational databases, we first note that data in a \C-set is organized into distinct tables, so queries over triangles of a semi-simplicial set do not have to consider vertices or edges, for example. Secondly, the uniqueness of foreign keys allows them to be indexed, which is crucial to performance when performing queries that require table joins. This mirrors the well-known performance differences between queries of data organized in relational databases versus knowledge graphs \cite{cheng2019category}. We compare both representations within the same rewriting tool in a single benchmark experiment, described in Figure \ref{fig:intbench}. This preliminary benchmark evaluates the performance of a single rewrite on semi-simplicial sets in a planar network of tessellated triangles. The rewrite locates a pair of triangles sharing an edge (i.e. a quadrilateral with an internal diagonal edge) and replaces them with a quadrilateral containing the opposite internal diagonal edge. We also chart the performance of finding all quadrilateral instances (homomorphisms) in variously sized grids. The results in Figure \ref{fig:intbench} demonstrate a lower memory footprint as well as improved rewrite and match searching for \C-sets.  

\section{Category-theoretic rewriting}
\subsubsection{Pushout complements}

Given a pair of arrows  ${A\xrightarrow{f}B\xrightarrow{g}C}$, one constructs a pushout {\it complement} by finding a pair of morphisms ${A\rightarrow D\rightarrow C}$ such that the resulting square is a pushout. While any category of \C-sets has pushouts, pushout complements are more subtle because they are not guaranteed to exist or be unique \cite{braatz2011delete}. These are both desirable properties to have when using the pushout complement in rewriting, so we will demand that identification  and dangling conditions (Eqs \ref{eq:t}-\ref{eq:u} \cite{lowe1993algebraic}) hold, which guarantee its existence, and that the first morphism, ${f: A\rightarrow B}$, be monic, which forces it to be unique. \cite{lack2005adhesive}

\begin{equation}
  \label{eq:t}
  \begin{gathered}
    \forall X \in \text{Ob}\ \mathcal{C}, \forall x_1, x_2 \in B_X: \\        
     g_X(x_1)=g_X(x_2) \implies x_1 = x_2 \lor \{x_1,x_2\}\subseteq f_X(A_X)
  \end{gathered}
\end{equation}

\begin{equation}
  \label{eq:u}
  \begin{gathered}
    \forall \phi: X\rightarrow Y \in \text{Hom}\ \mathcal{C}, \forall x \in C_X:\\        
    \phi(x) \in g_Y(B_Y - f_Y(A_Y)) \implies x \in g_X(B_X-  f_X(A_X))
  \end{gathered}
\end{equation}

\subsubsection{DPO, SPO, SqPO, PBPO+}
\begin{figure}[h!]
\centering
\includegraphics[width=\textwidth]{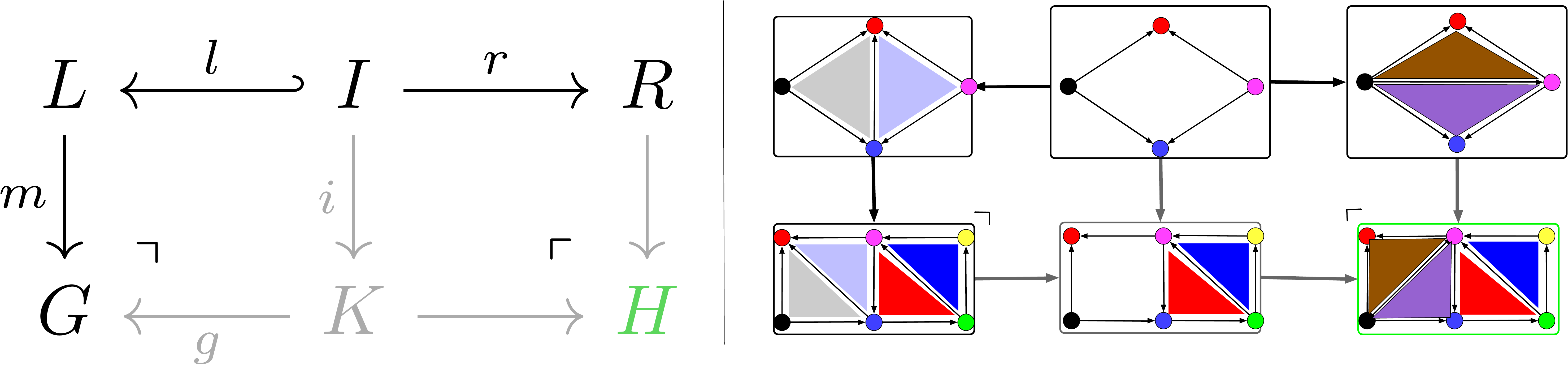}

\caption{{\bf Left: }DPO rewriting. Here and in the following figures, the initial data is in black, intermediate computations in grey, and the final result in green.  {\bf Right: } Application of a rewrite rule to flip the internal edge of a quadrilateral in a semi-simplicial set with two adjacent quadrilaterals. Here and in the following figures, colors are used to represent homomorphism data.}
\label{fig:dpo}
\end{figure}

The double-pushout (DPO) algorithm \cite{ehrig1973graph} formalizes a notion of rewriting a portion of a \C-set, visualized in Figure \ref{fig:dpo}.  The morphism $m$ is called the \textit{match} morphism. The meaning of $L$ is to provide a pattern that $m$ will match to a sub-\C-set in $G$, the target of rewriting. $R$ represents the \C-set which will be substituted back in for the matched pattern to yield the rewritten \C-set, and $I$ indicates what fragment of $L$ is preserved in the rewrite and its relation to $R$. To perform a rewrite, first, a pushout complement computes $K$, the original \C-set with deletions applied. Second, the final rewritten \C-set is computed via pushout along $r$ and $i$.

\begin{figure}[h!]
\centering
\includegraphics[width=.7\textwidth]{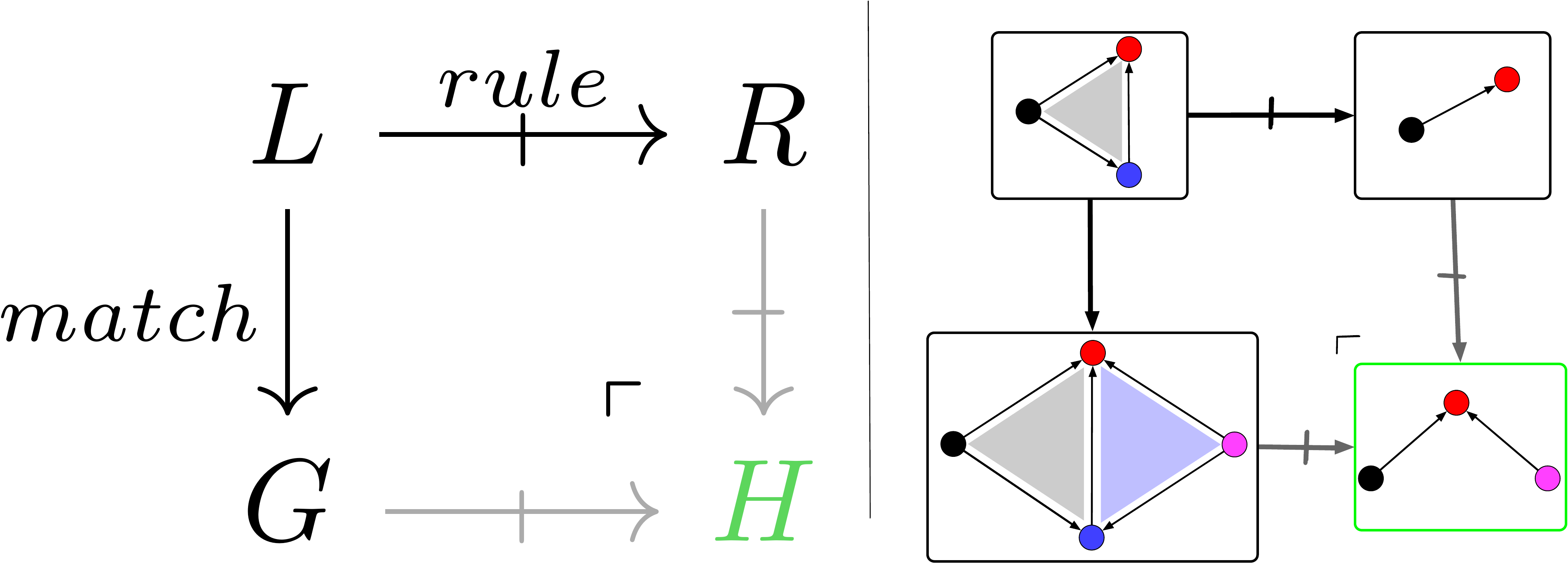}
\caption{{\bf Left: }SPO rewriting {\bf Right: } An instance of deletion in an unknown context.}
\label{fig:spo}
\end{figure}

\begin{figure}[h!]
\centering
\includegraphics[width=.8\textwidth]{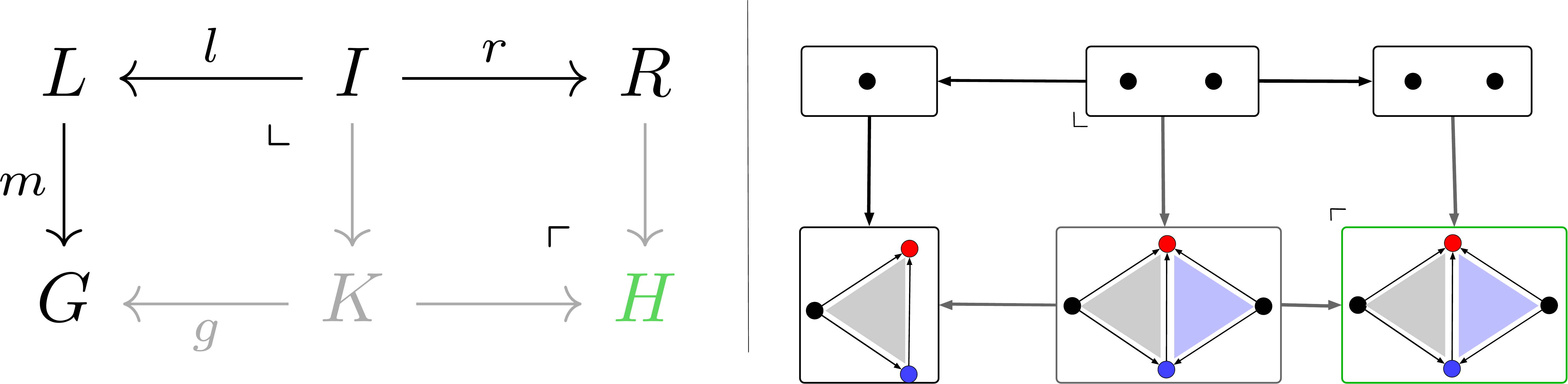}
\caption{{\bf Left: }SqPO rewriting {\bf Right: } an instance of creation in an unknown context. Note that there are multiple possible pushout complements because $l$ is not monic, but performing DPO using any of these would leave the original graph unchanged. Also note that enforcing the $\Delta_2$ equations (in Figure \ref{fig:d2}) when computing the partial object classifier affects the results: without equations, there are four resulting `triangle' objects, although two of these clearly do not form triangles.}
\label{fig:sqpo}
\end{figure}

\begin{figure}[h!]
\centering
\includegraphics[width=\textwidth]{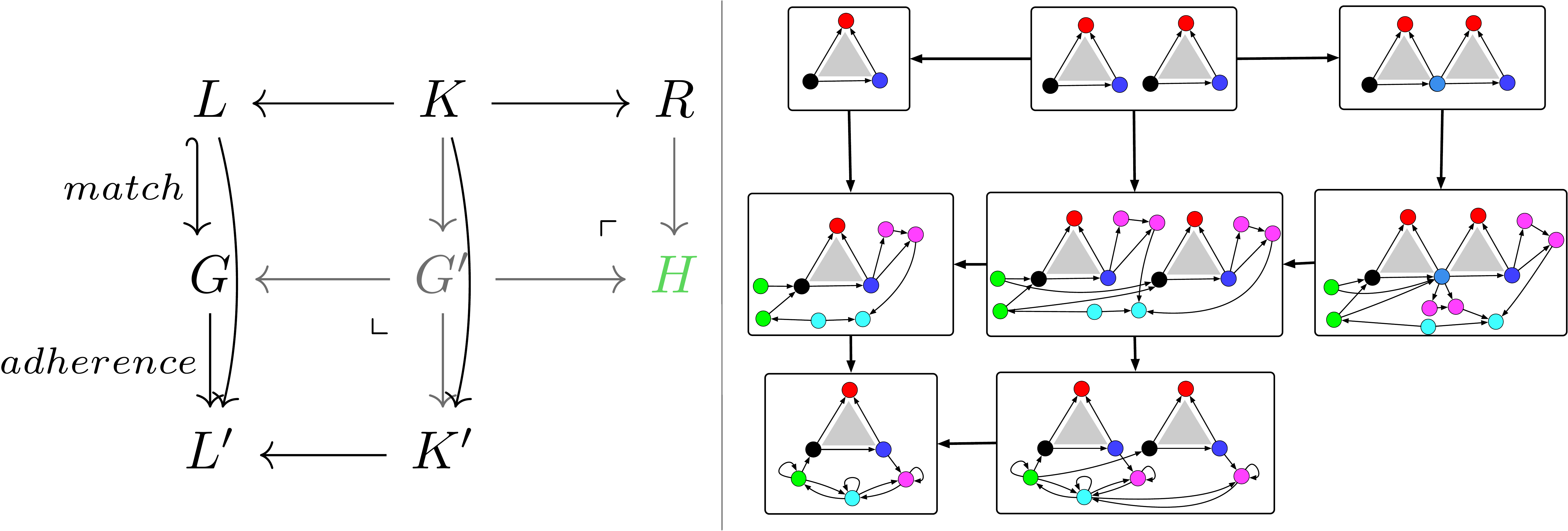}
\caption{{\bf Left: }PBPO+ rewriting {\bf Right: } an instance of rewriting where we explicitly control how the boundary of our matched triangular pattern is treated. The rule's type graph $L'$ says that, besides the matched pattern, we consider three other types of vertices: those that point at the black vertex (in green), those that are pointed at by the blue vertex (in pink) and the rest of the graph (light blue). The self loops on those extra vertices allow entire subgraphs to be mapped onto them, rather than just vertices. In $K'$, the rule indicates that we wish to duplicate the part of the graph that gets classified as pink (by the adherence map which assigns types to $G$), while only the {\it edges} from the green part of the graph will get copied when we duplicate the triangle. $L'$ has no notion of edges which are incident to the red vertex, so any input graph that has such an edge cannot be matched by this rule.}
\label{fig:pbpo}
\end{figure}

Single-pushout (SPO) rewriting  \cite{lowe1993algebraic} generalizes DPO rewriting, as every DPO transformation can be expressed as a SPO transformation. The additional expressivity allows us to delete in an unknown context, as demonstrated in Figure \ref{fig:spo}. The name comes from the construction being a single pushout in  the category of {\it partial} \C-set morphisms, \C-{\bf Par}. A partial \C-set morphism is a span $L \xhookleftarrow{l} I \xrightarrow{r} R$ where $l$ is monic. Sesqui-pushout (SqPO) rewriting \cite{corradini2006sesqui} is a more recent technique which generalizes the previous two. It is defined in terms of the notions of partial map classifiers and final pushout complements, and it further generalizes SPO by allowing both deletion and addition in an unknown context, as demonstrated in Figure \ref{fig:sqpo}. Lastly, Pullback-pushout+ (PBPO+) rewriting \cite{pbpo} is the most recent of the four paradigms we have implemented. As shown in Figure \ref{fig:pbpo}, each PBPO+ rule has its own type graph, $L^\prime$, which allows it to control rewriting of both the explicit matched pattern (described by $L$) as well as {\it all} elements in the input graph $G$ which interact with the boundary of the 
matched pattern. This means the notion of a match must be generalized from a match morphism $L\rightarrow G$ to include an adherence morphism $G \rightarrow L^\prime$ which is an interpretation of $G$ as typed over $L^\prime$.

\section{Design and implementation of generic categorical rewriting}

Within the paradigm of computational category theory, Catlab.jl is an open source framework for applied category theory at the center of an ecosystem of software packages called AlgebraicJulia \cite{patterson2021categorical,halter2020compositional}. We have recently added AlgebraicRewriting.jl to this ecosystem to support the categorical rewriting paradigms described above for \C-sets on finitely presented schemas \C. This class of structures balances expressivity and efficiency of manipulation, given that \C-sets are representable in the concrete language of relational databases \cite{schultz2016algebraic}, modulo equations in \C. In Catlab, each \C-set is automatically specialized to an efficient Julia data type; for example, when specialized to graphs, Catlab's implementation of \C-sets, performs competitively against libraries optimized for graphs \cite{patterson2021categorical}. Catlab now occupies a unique point in the space of rewriting software tools (Table 1). For performance in pattern matching (often the typical bottleneck of rewriting), Catlab outperforms ReGraph, the nearest alternative in terms of expressive capabilities (SqPO) and usability (Table 2).

\begin{table}[h!]
\centering
\begin{tabular}{lC{1cm}cC{1.2cm}C{.8cm}C{1cm}cC{1.4cm}C{1.3cm}c}
\toprule
Software & Typed Graphs & \C-sets & Rewrite type & CT Env & Last update &  GUI & Scripting\ \ Env & Library vs. App \\ \midrule
AGG\cite{taentzer2003agg}    & Y & N & S & N &  2017 & Y & N & Both\\ \midrule
Groove\cite{rensink2010user} & Y & N & S & N &  2021 & Y & N & App\\ \midrule
Kappa\cite{hayman2013pattern}  & N & N &  & N &  2021 & Y & Y & App\\ \midrule
VeriGraph\cite{azzi2018verigraph}  & Y & N & D & Y & 2017 & N & Y & Lib\\ \midrule
ReGraph\cite{harmer2020reversibility}  & Y & N & Q & N & 2018 & N & Y & Lib\\ \midrule
AlgebraicRewriting & Y & Y & D,S,Q,P & Y &  2022 & N & Y & Lib  \\ \bottomrule
\end{tabular}
\vspace{.5cm}
\caption{ High-level comparison with contemporary graph rewriting software packages. {\it Rewrite type} refers to whether DPO (D), SPO (S), SqPO (Q), and PBPO+ (P) are explicitly supported. {\it CT Env} refers to whether the software was implemented within a general environment of categorical abstractions beyond those immediately useful for graph rewriting. {\it Last update} refers to the year of the last minor version release (i.e. X.Y.0).} 
\label{tab:comp}
\end{table}

\begin{table}[h!]
\begin{minipage}{.4\textwidth}
\centering
\begin{tabular}{c|c|c}
    \toprule
     Mesh size & Catlab (s) & ReGraph (s) \\
     \midrule
     2 by 2 & $1.2\times 10^{-4}$ & $5.3\times 10^{-3}$  \\
     2 by 3 & $2.7\times 10^{-4}$ & 8.0 \\
     2 by 4 & $4.7\times 10^{-4}$ & 1313.3 \\
     2 by 5 &  $6.7\times 10^{-4}$ & 44979.8 \\
     \bottomrule
\end{tabular}
\label{tab:regraph-comp-table}

\end{minipage}
\hspace{0.12\textwidth}
\begin{minipage}{.5\textwidth}

\caption{Catlab $\mathcal{C}$-set homomorphism search compared to ReGraph typed graph homomorphism search. The task was to find all quadrilateral patterns in meshes of increasing size. Tests were conducted on a single AMD EPYC 75F3 Milan 3.0 GHz Core with 4GB of RAM. }
\end{minipage}
\end{table}

The development of Catlab has emphasized the separation of syntax and semantics when modeling a domain. This facilitates writing generic code, as diverse applications can share syntactic features, e.g. representability through string diagrams and hierarchical operad composition, with different semantic interpretations of that syntax for diverse applications. One result of this is that library code becomes very reusable, such that new features can be built from the composition of old parts with minimal additions, which reduces both developer time and the surface area for new bugs. 

This point is underscored by the developer experience of implementing the above rewriting algorithms: because limits and colimits already existed for \C-sets, PBPO+ required no serious code writing, and the implementation of DPO only required pushout complements. Like limits and colimits, pushout complements are computed component-wise for \C-sets, meaning that only basic code related to pushout complements of finite sets was required. More work was needed to implement SPO because no infrastructure for the category \C-{\bf Par} existed at the time. However, with a specification of partial morphism pushouts in terms of pushouts and pullback complements of total morphisms \cite[Theorem 3.2]{kennaway1990graph}, the only engineering required for this feature was an efficient pullback complement for \C-sets. Lastly, for SqPO, an algorithm for final pullback complements for \C-sets was the only nontrivial component that needed to be implemented, based on \cite[Theorem 1]{corradini2015agree} and \cite[Theorem 2]{behr2021concurrency}. This required generalizing examples of partial map classifiers from graphs to \C-sets. Because the partial map classifier can be infinite for even a finitely presented \C-set, this type of rewriting is restricted to acyclic schemas, which nevertheless includes graphs, Petri nets, semi-simplicial sets, and other useful examples.

Because AlgebraicJulia is a collection of libraries rather than a standalone application, users have a great deal of freedom in defining their own abstractions and automation techniques, using the full power of the Julia programming language. A great deal of convenience follows from having the scripting language and the implementation language be the same: we can specify the pattern of a rewrite rule via a pushout, or we can programmatically generate repetitive rewrite rules based on structural features of a particular graph. Providing libraries rather than standalone black-box software makes integration into other projects (in the same programming language) trivial, and in virtue of being open-source library, individuals can easily extend the functionality. By making these extensions publicly available, all members of the AlgebraicJulia ecosystem can mutually benefit from each other's efforts. As examples of this, the following additional features that have been contributed to AlgebraicRewriting.jl all serve to extend its utility as a general rewriting tool:

\subsection{Computation of homomorphisms and isomorphisms of C-sets}
For rewriting algorithms to be of practical use, morphisms matching the left-hand-side of rules must somehow be supplied. The specification of a \C-set morphism requires a nontrivial amount of data that must satisfy the naturality condition. Furthermore, in confluent rewriting systems, manually finding matches is an unreasonable request to make of the end user, as the goal is to apply all rewrites possible until the term reaches a normal form. For this reason, DPO rewriting of \C-sets benefits from a generic algorithm to find homomorphisms, analogous to structural pattern matching in the tree term rewriting case.

The problem of finding a \C-set homomorphism $X \to Y$, given a finitely presented category \C~and two finite \C-sets $X$ and $Y$, is generically at least as hard as the graph homomorphism problem, which is NP-complete. On the other hand, the \C-set homomorphism problem can be framed as a constraint satisfaction problem (CSP), a classic problem in computer science for which many algorithms are known \cite[Chapter 6] {russell2010ai}. Since \C-sets are a mathematical model of relational databases \cite{spivak2012functorial}, the connection between \C-set homomorphisms and constraint satisfaction is a facet of the better-known connection between databases and CSPs \cite{vardi2000constraint}.

To make this connection precise, we introduce the slightly nonstandard notion of a typed CSP. Given a finite set $T$ of \emph{types}, the slice category $\mathbf{FinSet}/T$ is the category of \emph{$T$-typed finite sets}. A \emph{typed CSP} then consists of $T$-typed finite sets $V$ and $D$, called the \emph{variables} and the \emph{domain}, and a finite set of \emph{constraints} of form $(\mathbf{x}, R)$, where $\mathbf{x} = (x_1,\dots,x_k)$ is a list of variables and $R \subseteq D^{-1}(V(x_1)) \times \cdots \times D^{-1}(V(x_k))$ is a compatibly typed $k$-ary relation. An \emph{assignment} is a map $\phi: V \to D$ in $\mathbf{FinSet}/T$. The objective is to find a \emph{solution} to the CSP, namely an assignment $\phi$ such that $(\phi(x_1),\dots,\phi(x_k)) \in R$ for every constraint $(\mathbf{x}, R)$.

The problem of finding a \C-set morphism $X \to Y$ translates to a typed CSP by taking the elements of $X$ and $Y$ to be the variables and the domain of the CSP, respectively. To be precise, let the types $T$ be the objects of \C. The variables $V: \{(c,x): c \in \mathcal{C}, x \in X(c)\} \to \Ob 
\mathcal{C}$ are given by applying the objects functor $\Ob: \mathbf{Cat} \to \mathbf{Set}$ to $\int X \to \mathcal{C}$, the category of elements of $X$ with its canonical projection. Similarly, the domain is $D := \Ob(\int Y \to \mathcal{C})$. Finally, for every generating morphism $f: c \to c'$ of \C~and every element $x \in X(c)$, introduce a constraint $((x,x'),R)$ where $x' := X(f)(x)$ and $R := \{(y,y') \in Y(c) \times Y(c'): Y(f)(y) = y'\}$ is the graph of $Y(f)$. By construction, an assignment $\phi: V \to D$ is the data of a \C-set transformation (not necessarily natural) and $\phi$ is a solution if and only if the transformation is natural. Thus, the solutions of the typed CSP are exactly the \C-set homomorphisms $ X \to Y$.

With this reduction, CSP algorithms are straightforwardly ported to algorithms for finding \C-set morphisms, where the types and special structure permits optimizations, one example being the use of the discrete opfibration condition to accelerate the search. We only consider assignments that satisfy the typing relations. We have adapted backtracking search \cite[Section 6.3]{russell2010ai}, a simple but fundamental CSP algorithm, to find \C-set homomorphisms. By also maintaining a partial inverse assignment, this algorithm is easily extended to finding \C-set monomorphisms, an important constraint when matching for rewriting. Since a monomorphism between finite \C-sets $X$ and $Y$ is an isomorphism if and only if $X(c)$ and $Y(c)$ have the same cardinality for all $c \in$ \C, this extension also yields an algorithm for isomorphism testing, which is useful for checking the correctness of rewrites.

\subsection{Diagrammatic syntax}
\begin{figure}
\centering
\begin{subfigure}{\textwidth}
\begin{equation*}
\begin{tikzcd}
	&&& {\texttt{Entity}} \\
	& {\texttt{Food}} &&&& {\texttt{Container}} \\
	{\texttt{Egg}} && {\texttt{YolkWhite}} && {\texttt{Bowl}} && {\texttt{Table}}
	\arrow["{\texttt{food\_is\_entity}}"{description}, from=2-2, to=1-4]
	\arrow["{\texttt{container\_is\_entity}}"{description}, from=2-6, to=1-4]
	\arrow["{\texttt{food\_in\_on}}"{description}, from=2-2, to=2-6]
	\arrow["{\texttt{bowl\_is\_container}}"{description}, from=3-5, to=2-6]
	\arrow["{\texttt{egg\_is\_food}}"{description}, from=3-1, to=2-2]
	\arrow["{\texttt{yolk\_white\_is\_food}}"{description}, from=3-3, to=2-2]
	\arrow["{\texttt{table\_is\_container}}"{description}, from=3-7, to=2-6]
\end{tikzcd}
\end{equation*}
\caption{Fragment of a schema that models recipes for cooking breakfast}
\end{subfigure}
\par\bigskip
\begin{subfigure}{\textwidth}
\begin{center}
\includegraphics[width=.8\textwidth]{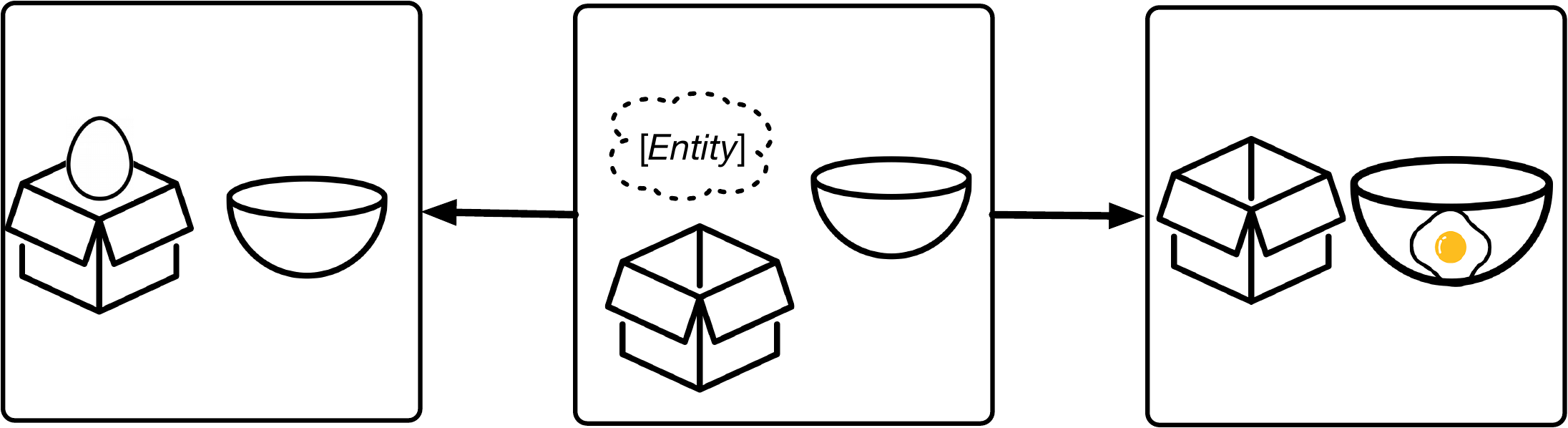}
\end{center}

\caption{Cartoon visualization of egg cracking rule. Notably we require an abstract entity in the interface, mapping to both the egg and yolk+white, to reflect that they are the same entity.}
\end{subfigure}
\par\bigskip

\begin{subfigure}{\textwidth}
\begin{minted}[fontsize=\footnotesize]{julia}
crack_egg_in_bowl = @migration SchCospan SchBreakfastKitchen begin
  L => @join begin # left-hand side of rule
    bowl::Bowl
    egg::Egg
  end
  I => @join begin # intermediate state of rule
    bowl::Bowl
    egg_entity::Entity       # entity underlying egg and yolk-white
    old_container::Container # original container of food
  end
  R => @join begin # right-hand side of rule
    bowl::Bowl
    yolk_white::YolkWhite
    food_in_on(yolk_white_is_food(yolk_white)) == bowl_is_container(bowl)
    old_container::Container
  end
  l => begin # left map in rule
    bowl => bowl
    egg_entity => food_is_entity(egg_is_food(egg))
    old_container => food_in_on(egg_is_food(egg))
  end
  r => begin # right map in rule
    bowl => bowl
    egg_entity => food_is_entity(yolk_white_is_food(yolk_white))
    old_container => old_container
  end
end
\end{minted}
\caption{DPO rewrite rule specified using diagrammatic syntax. This syntax allows us to avoid explicitly treating the underlying entity of the container, for example.}
\end{subfigure}

\caption{Example of a DPO rewrite rule specified using the diagrammatic syntax, adapted from a planning system for the cooking domain.}
\label{fig:diagrammatic-syntax}
\end{figure}

Specifying DPO rewrite rules can be cumbersome as a significant amount of combinatorial data is contained in a span of \C-sets. To make our system more user-friendly, we have developed a symbolic domain-specific language (DSL) to specify rewrite rules, based on the idea of assembling \C-sets from the atomic ones known as \emph{representables}. This involves no loss of generality since every \C-set can be expressed as a colimit of representable \C-sets \cite[Theorem 6.5.7]{riehl2016}. For instance, in the category of graphs, the two representables are the graphs with one isolated vertex and with one edge between two distinct vertices, and clearly every graph is a colimit of copies of these two graphs. An example of specifying a rewrite rule in this manner, using a much more elaborate schema, is shown in Figure \ref{fig:diagrammatic-syntax}.

The mathematics behind our DSL uses the underappreciated fact that the diagrams in a given category are themselves the objects of a category; as described in \cite{peschke2020diagrams,perrone2022,patterson2022diagrams} and references therein. Given a category $\cat{S}$, the \emph{diagram category} $\Diag(\cat{S})$ has, as objects, diagrams ${D: \cat{J} \to \cat{S}}$ in $\cat{S}$, and as morphisms $(\cat{J},D) \to (\cat{J}', D')$, a functor ${R: \cat{J} \to \cat{J}'}$ along with a natural transformation $\rho: D \Rightarrow D' \circ R$. Another diagram category $\Diag^{\co}(\cat{S})$ is defined similarly, except that the natural transformation in a morphism $(R,\rho)$ goes in the opposite direction: $\rho: D' \circ R \Rightarrow D$.

We now show that a span in $\Diag^{\co}(\cat{C})$ presents a span in $\cat{C}\text{-}\Set$, i.e., a DPO rewrite rule for \C-sets, as colimits of representables and morphisms between them. The category $\Diag^{\co}(\cat{C})$ has the advantage of referring only to the schema $\cat{C}$ and so can be described syntactically given a finite presentation of $\cat{C}$.

\begin{proposition}
By applying the Yoneda embedding and taking colimits, a span in the category $\Diag^{\co}(\cat{C})$ induces a span of $\cat{C}$-sets.
\end{proposition}
\begin{proof}
It is enough to define a functor $\Diag^{\co}(\cat{C}) \to \cat{C}\text{-}\Set$, which we do as the following composite
\begin{equation*}
\Diag^{\co}(\cat{C})
  \xrightarrow{\op} \Diag(\cat{C}^{\op})
  \xrightarrow{\Diag(y)} \Diag(\cat{C}\text{-}\Set)
  \xrightarrow{\colim} \cat{C}\text{-}\Set,
\end{equation*}
where $\op: \mathbf{Cat}^{\co} \to \mathbf{Cat}$ is the oppositization 2-functor and $y: \cat{C}^{\op} \to \cat{C}\text{-}\Set$ is the Yoneda embedding for $\cat{C}$. We are using the facts that the diagram construction extends to a (2-)functor $\Diag: \mathbf{Cat} \to \mathbf{Cat}$ in which morphisms act by postcomposition \cite[\S 2.1]{perrone2022} and that taking colimits is functorial with respect to the category $\Diag(\cat{S})$ whenever $\cat{S}$ is cocomplete \cite[\S 5.1]{perrone2022}.
\end{proof}

\subsection{Typed graph rewriting with slice categories}
Slice categories offer a form of constraining \C-sets without altering the schema. Consider the example of rewriting string diagrams encoded as hypergraph cospans \cite{bonchi2020string}. These can be used to represent terms in a symmetric monoidal theory, where it is important to restrict diagrams to only those which draw from a fixed set of boxes with particular arities, given by a monoidal signature $\Sigma$, which induces the unique hypergraph $H\Sigma$ which has all box types from $\Sigma$ and a single vertex. Working within the slice category $\mathbf{Hyp}/H\Sigma$ prevents us from performing rewrites which violate the arities of the operations specified by $\Sigma$.

There are two ways to implement rewriting in \C{\bf-Set}$/X$ for a particular \C: the computation can be performed with the objects $L, I, R, G$ being \C-set morphisms, or it can be performed in $[\int X, \mathbf{Set}]$. Programming with generic categorical abstraction greatly lowered the barrier to implementing both of these: for the former, what was needed was to relate the pushout and pushout complement of \C{\bf-Set}$/X$ to the corresponding computations in \C{\bf-Set}. The barrier to the latter was to compute the category of elements and migrate data between the two representations, code which had already been implemented. As the former strategy requires less data transformation, it is preferred. 

\subsection{Open system rewriting with structured cospans}
The forms of rewriting discussed up to this point have concerned rewriting closed systems. Structured cospans are a general model for open systems, which formalize the notion of gluing together systems which have designated inputs and outputs. Open systems are modeled as cospans of form $La \rightarrow x \leftarrow Lb$, where the apex $x$ represents the system itself and the feet $La$ and $Lb$ represent the inputs and outputs, typically discrete systems such as graphs without edges. Here, $L: A \rightarrow X$ is a functor  that maps from the system category $A$ to the system interface category $X$, and $L$ must be a left adjoint between categories with finite colimits.\footnote{The $L$ of structured cospans should not be confused with the $L$ of the rewrite rule $L\leftarrow I \rightarrow R$.} Larger systems are built up from smaller systems via pushouts in $X$, which glue systems together along a shared interface: $(La\rightarrow x \leftarrow Lb \rightarrow y \leftarrow Lc) \mapsto (La \rightarrow x+_{Lb}y \leftarrow Lc)$.

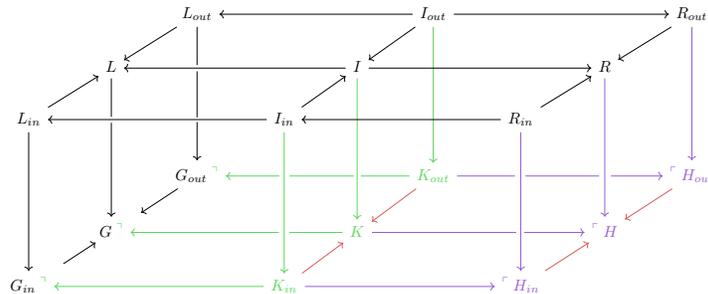
\begin{figure}[h!]
\centering
\adjustbox{scale=0.6,center}{%
\begin{tikzcd}
	&& {L_{out}} &&& {I_{out}} &&& {R_{out}} \\
	& L &&& I &&& R \\
	{L_{in}} &&& {I_{in}} &&& {R_{in}} \\
	&& {G_{out}\ \textcolor{rgb,255:red,92;green,214;blue,92}{^{\urcorner}}} &&& \textcolor{rgb,255:red,92;green,214;blue,92}{K_{out}} &&& \textcolor{rgb,255:red,153;green,92;blue,214}{^{\ulcorner}\ H_{out}} \\
	& {G \ \textcolor{rgb,255:red,92;green,214;blue,92}{^{\urcorner}}} &&& \textcolor{rgb,255:red,92;green,214;blue,92}{K} &&& \textcolor{rgb,255:red,153;green,92;blue,214}{^{\ulcorner}\ H} \\
	{G_{in}\ \textcolor{rgb,255:red,92;green,214;blue,92}{^{\urcorner}}} &&& \textcolor{rgb,255:red,92;green,214;blue,92}{K_{in}} &&& \textcolor{rgb,255:red,153;green,92;blue,214}{^{\ulcorner}\ H_{in}}
	\arrow[from=3-1, to=6-1,]
	\arrow[from=1-3, to=4-3, shorten >=60pt, no head]
	\arrow[from=1-3, to=4-3, shorten <=28pt, shorten >=27pt, no head]
	\arrow[from=1-3, to=4-3, shorten <=60pt]
	\arrow[draw={rgb,255:red,153;green,92;blue,214}, from=1-9, to=4-9]
	\arrow[draw={rgb,255:red,153;green,92;blue,214}, from=2-8, to=5-8]
	\arrow[draw={rgb,255:red,153;green,92;blue,214}, from=3-7, to=6-7]
	\arrow[draw={rgb,255:red,92;green,214;blue,92}, from=1-6, to=4-6]
	\arrow[draw={rgb,255:red,92;green,214;blue,92}, from=2-5, to=5-5]
	\arrow[draw={rgb,255:red,92;green,214;blue,92}, from=3-4, to=6-4]
	\arrow[draw={rgb,255:red,92;green,214;blue,92}, from=6-4, to=6-1]
	\arrow[draw={rgb,255:red,153;green,92;blue,214}, from=6-4, to=6-7]
	\arrow[draw={rgb,255:red,92;green,214;blue,92},  shorten <=82pt, from=4-6, to=4-3]
	\arrow[draw={rgb,255:red,92;green,214;blue,92},  shorten >=85pt, no head, from=4-6, to=4-3]
	\arrow[draw={rgb,255:red,92;green,214;blue,92},  shorten <=35pt, shorten >=40pt, no head, from=4-6, to=4-3]
	\arrow[draw={rgb,255:red,214;green,92;blue,92}, from=6-7, to=5-8]
	\arrow[draw={rgb,255:red,214;green,92;blue,92}, from=4-9, to=5-8]
	\arrow[draw={rgb,255:red,214;green,92;blue,92}, from=4-6, to=5-5]
	\arrow[from=3-1, to=2-2]
	\arrow[from=1-3, to=2-2]
	\arrow[from=1-6, to=2-5]
	\arrow[from=3-4, to=2-5]
	\arrow[from=3-7, to=2-8]
	\arrow[from=1-9, to=2-8]
	\arrow[draw={rgb,255:red,214;green,92;blue,92}, from=6-4, to=5-5]
	\arrow[from=2-5, to=2-2]
	\arrow[from=2-5, to=2-8]
	\arrow[from=1-6, to=1-3]
	\arrow[from=1-6, to=1-9]
	\arrow[from=2-2, to=5-2, shorten >=63pt, no head]
	\arrow[from=2-2, to=5-2, shorten <=28pt]
	\arrow[draw={rgb,255:red,92;green,214;blue,92}, shorten <=40pt, from=5-5, to=5-2]
	\arrow[draw={rgb,255:red,92;green,214;blue,92}, shorten >=100pt, no head, from=5-5, to=5-2]
	\arrow[shorten >=8pt, from=4-3, to=5-2]
	\arrow[shorten <=8pt, from=6-1, to=5-2]
	\arrow[draw={rgb,255:red,153;green,92;blue,214},  shorten <=96pt, from=5-5, to=5-8]
	\arrow[draw={rgb,255:red,153;green,92;blue,214},  shorten >=43pt, no head, from=5-5, to=5-8]
	\arrow[from=3-7, to=3-4]
	\arrow[from=3-4, to=3-1]
	\arrow[draw={rgb,255:red,153;green,92;blue,214}, shorten <=97pt, from=4-6, to=4-9]
	\arrow[draw={rgb,255:red,153;green,92;blue,214}, shorten >=93pt, no head,from=4-6, to=4-9]
	\arrow[draw={rgb,255:red,153;green,92;blue,214}, shorten <=43pt, shorten >=40pt, no head,from=4-6, to=4-9]
\end{tikzcd}
}
\caption{Applying a structured cospan rewrite rule. \C-sets and morphisms in black are the initial data: the upper face represents the open rewrite rule, the upper left edge represents the open pattern to be matched, and the left face represents the matching. Green morphisms are computed by pushout complement in \C-$\mathbf{Set}$. The purple morphisms are computed by the rewriting pushouts and red morphisms are computed by the structured cospan pushouts. Figure adapted from \cite[Section 4.2]{cicala2019rewriting}.}
\label{fig:openrewrite}
\end{figure}

\begin{figure}[h!]
\centering
\includegraphics[width=.8\textwidth]{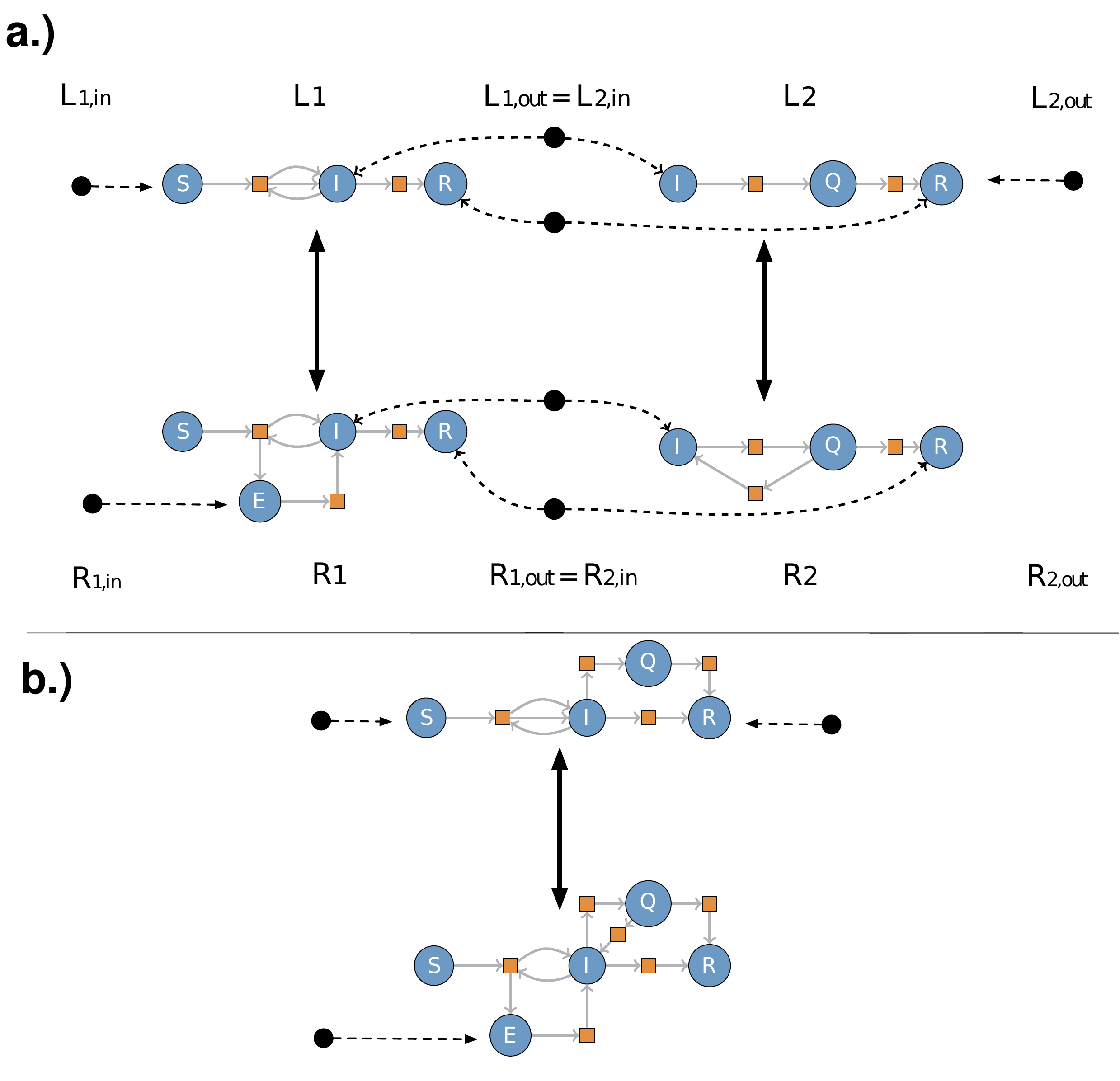}
\caption{{\bf a.)} Example of horizontal composition of structured cospan rewrite rules. The $L$ and $R$ structured cospans are positioned on the top and bottom, respectively. For clarity, $I$ cospans are omitted. {\bf b.)} The result of composition. } 
\label{fig:openp}
\end{figure}

When $L$, $I$, and $R$ are each structured cospans, there is extra data to consider when rewriting, as shown in Figure \ref{fig:openrewrite}.
In ordinary DPO rewriting, if the $R$ of one rewrite rule equals the $L$ of another, a composite rewrite rule can be constructed, which could be called \emph{vertical} composition. In the case of structured cospans, \emph{horizontal} composition emerges from composing the $L$, $I$, and $R$ of two structured cospan rules pairwise, visualized in Figure \ref{fig:openp}. These two forms of composition together yield a double category of structured cospan rewrites, where horizontal arrows are in correspondence with structured cospans and squares are in correspondence with all possible rewrites \cite{cicala2019rewriting}. %

While this compositional approach to building open systems can be an illuminating way to organize information about a complex system, there can also be computational benefits. When searching for a match in a large \C-set, the search space grows as $O(n^k)$ where $k$ is the size of the pattern $L$ and $n$ is the size of $G$. However, after decomposing $G$ into a composite of substructures and restricting matches to homomorphisms into a specific substructure, the search space is limited by $O(m^k)$ where $m<n$ is the size of the substructure. Not only does this accelerate the computation, but it can be semantically meaningful to restrict matches to those which do not cross borders. 

\subsection{Distributed graph rewriting} 
\begin{figure}[h!]
\centering
\includegraphics[width=.8\textwidth]{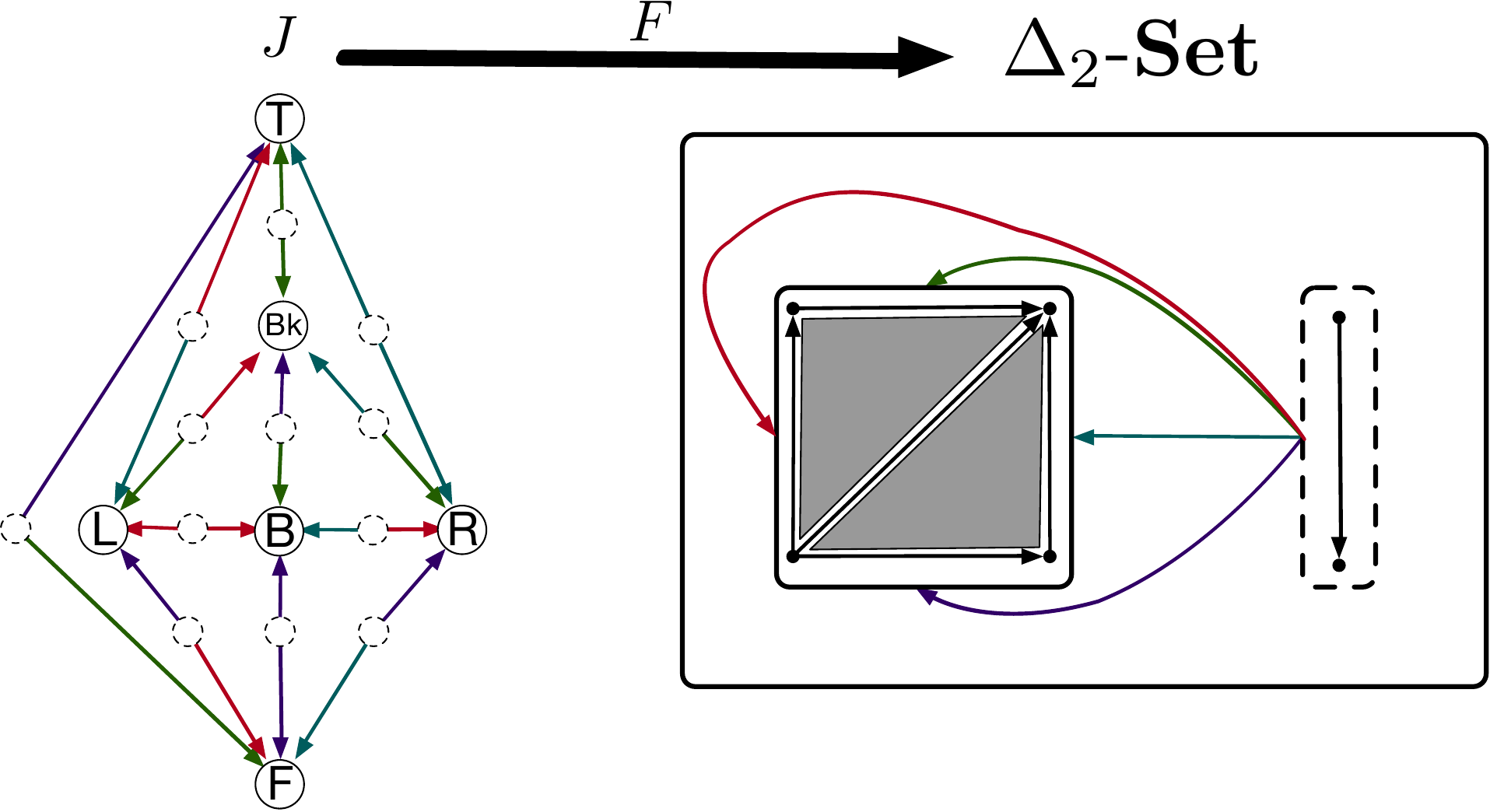}
\caption{Constructing the surface of a cube compositionally with a distributed graph. $F$ sends the solid circles to the square face graph and the dashed circles to the edge graph. Colors indicate which morphism from the edge to the face which controls how the faces are being glued together. We construct the assembled cube as a \C-set simply by taking the colimit of the diagram.}
\label{fig:dist}
\end{figure}

Distributed graphs offer an alternative formalism that allows one to decompose a large graph into smaller ones while maintaining consistency at the boundaries, and thus it is another strategy for parallelizing computations over graphs. The content of a distributed graph can be succinctly expressed in the language of category theory as a diagram in {\bf Grph}. Because Catlab has sophisticated infrastructure in place for manipulating categories of diagrams, it merely takes specializing the codomain of the Diagram datatype to {\bf Grph} to represent distributed graphs and their morphisms. Note that we can easily generalize to distributed semi-simplicial sets or other \C-sets (Figure \ref{fig:dist}). Colimits in the category of diagrams (in a cocomplete category) are defined in terms of left Kan extensions \cite{peschke2020diagrams}, and with our implementation \cite{modexplore} it is possible to develop a rewriting tool for distributed graphs.

\subsection{Graph processes}
\begin{figure}[h!]
\centering

\[\begin{tikzcd}
	{L_1} & {I_1} & {R_1\ L_2} & {I_2} & {R_2\ ...} \\
	\textcolor{rgb,255:red,214;green,92;blue,92}{G_1} & \textcolor{rgb,255:red,214;green,92;blue,92}{K_1} & \textcolor{rgb,255:red,214;green,92;blue,92}{G_2} & \textcolor{rgb,255:red,214;green,92;blue,92}{K_2} & \textcolor{rgb,255:red,214;green,92;blue,92}{G_3\ ...} \\
	&& {\Sigma G}
	\arrow[draw={rgb,255:red,214;green,92;blue,92}, from=2-2, to=2-1]
	\arrow[shift right=1, draw={rgb,255:red,214;green,92;blue,92}, from=2-2, to=2-3]
	\arrow[shift left=1, draw={rgb,255:red,214;green,92;blue,92}, from=2-4, to=2-3]
	\arrow[shift right=1, draw={rgb,255:red,214;green,92;blue,92}, from=2-4, to=2-5]
	\arrow["{c_1}"', shift right=2, from=1-3, to=2-3]
	\arrow["{m_2}", shift left=2, from=1-3, to=2-3]
	\arrow[from=1-4, to=1-3]
	\arrow[from=1-2, to=1-3]
	\arrow[from=1-2, to=1-1]
	\arrow[from=1-4, to=1-5]
	\arrow[""{name=0, anchor=center, inner sep=0}, "{m_1}"', from=1-1, to=2-1]
	\arrow[from=1-4, to=2-4]
	\arrow["{\iota_1}"', shift right=2, tail, from=2-1, to=3-3]
	\arrow["{\iota_2}"', tail, from=2-3, to=3-3]
	\arrow["{\iota_3}", shift left=2, tail, from=2-5, to=3-3]
	\arrow[from=1-2, to=2-2]
	\arrow["{c_2}", shift right=2, from=1-5, to=2-5]
	\arrow["\lrcorner"{anchor=center, pos=0.125, rotate=90}, shift right=1, draw=none, from=2-3, to=1-4]
	\arrow["\lrcorner"{anchor=center, pos=0.125, rotate=180}, shift left=2, draw=none, from=2-5, to=1-4]
	\arrow["\lrcorner"{anchor=center, pos=0.125, rotate=180}, shift left=1, draw=none, from=2-3, to=1-2]
	\arrow["\lrcorner"{anchor=center, pos=0.125, rotate=90}, draw=none, from=2-1, to=1-2]
\end{tikzcd}\]

\caption{The graph processes construction from a sequence of rewrites with match morphisms $m_i$ and co-match morphisms $c_i$ labeled. $\Sigma G$ is constructed as the colimit of the red subdiagram, and its role is to identify the same elements across time, if we interpret $G_i$ as a temporal sequence. Therefore, given a notion of element production, deletion, and preservation, if $i$ produces some element that $j$ preserves or deletes, there must be a causal dependency $i < j$.}
\label{fig:proc}
\end{figure}
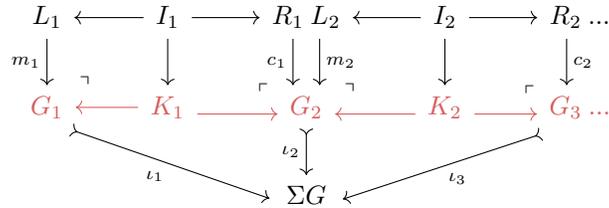

Given a concrete sequence of rewrites, perhaps representing a sequence of actions required to take one from an initial state to some desired state, it is of practical importance to represent the steps taken in a maximally-parallel manner that has only the necessary dependencies, such as one rewrite step creating an element that another rewrite step deletes. Graph processes \cite{corradini1996graph} are a construction which exposes the causal dependencies between rewrites as a partially-ordered set. The construction of this partial order is expressed as a colimit of a certain bipartite diagram, as shown in Figure \ref{fig:proc}. Colimits of diagrams being readily computable in Catlab led to this extension requiring only a small amount of programmer effort.

\subsection{Further extensions}
Examples of further features, such as negative application conditions, parallel rewriting, rewriting with functions applied to attributes, matching variables on attributes, (e.g. one rule which can identify any triangle that has exactly two edges with an equal length attribute and rewrite to make all three edges have that length) are found in AlgebraicRewriting documentation or tests.

\section{Conclusions and Future Work}

There are many desiderata for software development in academic and industrial settings alike, such as velocity of development, robustness to future changes in design, and correctness. We demonstrated how designing software with category-theoretic abstractions facilitates the achievement all three of these, using the mature field of graph rewriting software as a case study.  

While current graph transformation software in use is often very specialized to particular domains, such as chemistry, we show that DPO, SPO, SqPO, and PBPO+ rewriting can be efficiently performed on \C-sets, which are viewed as a subset of typed graphs (discrete opfibrations) with desirable theoretical and performance characteristics, and we have presented the first practical implementation for this. This result allows generic rewrite operations to be used in a variety of contexts, when it would otherwise be time-consuming and error-prone to develop custom rewrite algorithms for such a multitude of data structures or to work with typed graphs and enforce the discrete opfibration condition by other means.  We also extended these implementations to the first practical implementations of homomorphism search, structured cospan rewriting, and distributed graphs for arbitrary \C-sets. Our internal benchmark showed that \C-set rewriting can leverage the discrete opfibration condition to outperform typed graphs in memory and speed, and an external benchmark showed a significant speedup relative to comparable graph rewriting software.

Catlab and AlgebraicRewriting could be extended to a tool for graph transformation researchers to computationally validate and explore new ideas. Researchers interested developing tools to be directly consumed by others could produce a performant and easily interoperable instantiation of their work. Even those interested in rewriting systems as mathematical objects can benefit from this process by gaining intuition and empirically testing conjectures about their constructions. However, many useful concepts from graph rewriting have yet to be added, such as rule control mechanisms and rule algebras, but the extensibility of Catlab allows researchers to do this on their own or with the support of Catlab's active user community. 

To create tools for practicing scientists and engineers, our future work involves building practical scientific software that applies rewriting in each its main areas, i.e. graph {\it relations}, {\it languages}, and {\it transition systems}: respectively, a theorem prover for symmetric monoidal categories by performing e-graph equality saturation \cite{willsey2021egg} with rewriting, a tool for defining and exploring a language of open epidemiological models, and a general agent-based model simulator.

\bibliographystyle{splncs04}
\bibliography{./references}
\end{document}